\documentclass[journal]{IEEEtran}
\usepackage[font=small,skip=1pt]{caption}
\IEEEoverridecommandlockouts
\usepackage{textcomp}
\usepackage{gensymb}
\usepackage{latexsym}
\usepackage[T1]{fontenc}
\usepackage{multirow}
\usepackage{graphicx}
\usepackage{varioref}
\usepackage{array}
\usepackage{enumitem}
\usepackage{soul,color}
\usepackage{longtable}
\newcommand{\cmark}{\ding{51}}%
\newcommand{\xmark}{\ding{55}}%
\usepackage{algorithmic}
\usepackage[ruled, vlined]{algorithm2e}
\usepackage{pifont}
\usepackage{booktabs}

\ifCLASSOPTIONcompsoc
  \usepackage[caption=false,font=normalsize,labelfont=sf,textfont=sf]{subfig}
\else
  \usepackage[caption=false,font=footnotesize]{subfig}
\fi

\usepackage{color}
\definecolor{light}{rgb}{0.5, 0.5, 0.5}

\usepackage{graphicx}
\medskip
\usepackage{amssymb}
\usepackage{amsmath,epsfig,amssymb}
\usepackage{amsthm}
\usepackage{xcolor}
\usepackage[active]{srcltx} 
\usepackage{epstopdf} 

\usepackage{bm}
\usepackage{amsthm}
\usepackage{enumitem}
\DeclareRobustCommand*{\IEEEauthorrefmark}[1]{%
  \raisebox{0pt}[0pt][0pt]{\textsuperscript{\footnotesize\ensuremath{#1}}}}

\usepackage{cite}
\usepackage{amsmath,amssymb,amsfonts}
\usepackage{algorithmic}
\usepackage{graphicx}
\usepackage{textcomp}
\usepackage{xcolor}
\def\BibTeX{{\rm B\kern-.05em{\sc i\kern-.025em b}\kern-.08em
    T\kern-.1667em\lower.7ex\hbox{E}\kern-.125emX}}
\usepackage{lineno,hyperref}
\usepackage[utf8]{inputenc}
\usepackage[english]{babel}
\usepackage{mathdots}

\graphicspath{{Pictures/}{../jpeg/}} 
\usepackage{float}
\usepackage{amsmath}

\begin{document}

\title{Unveiling Thoughts: A Review of Advancements in EEG Brain Signal Decoding into Text}

\author{\IEEEauthorblockN{Saydul Akbar Murad\IEEEauthorrefmark{1}, ~\IEEEmembership{(IEEE Member)}
Nick Rahimi\IEEEauthorrefmark{1},~\IEEEmembership{(IEEE Member)}}

\thanks{\IEEEauthorrefmark{1} School of Computing Sciences \& Computer Engineering, University of Southern Mississippi, Hattiesburg, MS, USA. (e-mail: saydulakbar.murad@usm.edu, nick.rahimi@usm.edu)}
}  




\maketitle
\begin{abstract}
The conversion of brain activity into text using electroencephalography (EEG) has gained significant traction in recent years. Many researchers are working to develop new models to decode EEG signals into text form. Although this area has shown promising developments, it still faces numerous challenges that necessitate further improvement.  It's important to outline this area's recent developments and future research directions. In this review article, we thoroughly summarize the progress in EEG-to-text conversion. Firstly, we talk about how EEG-to-text technology has grown and what problems we still face. Secondly, we discuss existing techniques used in this field. This includes methods for collecting EEG data, the steps to process these signals, and the development of systems capable of translating these signals into coherent text. We conclude with potential future research directions, emphasizing the need for enhanced accuracy, reduced system constraints, and the exploration of novel applications across varied sectors. By addressing these aspects, this review aims to contribute to developing more accessible and effective Brain-Computer Interface (BCI) technology for a broader user base.
\end{abstract} 

\section{Introduction}
EEG-based brain-to-text communication is a revolutionary field that explores using EEG to decode brain signals into actual text. EEG measures electrical activity on the scalp, and researchers are currently devising techniques to convert particular patterns into letters, words, and even sentences. This technological advancement exhibits significant promise for persons who experience speech or motor disabilities, offering a direct communication channel that bypasses traditional methods. BCIs play a pivotal role in EEG-based brain-to-text communication. BCIs bridge the gap between the brain and external devices by interpreting brain signals to facilitate control or communication. They have evolved significantly, from early systems focusing on simple commands to more advanced interfaces capable of recognizing complex thought patterns. This evolution is important for brain-to-text communication, as researchers rely on BCIs to translate the intricate neural correlates of language into meaningful text.

In recent years, there has been a significant increase in research focused on decoding EEG signals for the purpose of direct brain-to-text communication \cite{intro1}. The increasing interest can be attributed to various things. Firstly, the progress in machine learning algorithms facilitated the researchers to analyze complex EEG patterns with greater accuracy \cite{intro2}. Secondly, the potential applications of this technology are vast, particularly in the realm of communication assistance \cite{intro3}. Finally, the non-invasive nature of EEG makes it a promising option for individuals who are unable to employ conventional modes of communication. For individuals with conditions like amyotrophic lateral sclerosis (ALS), stroke, or severe cerebral palsy, traditional communication methods can be severely limited or even impossible \cite{intro1}.  Brain-to-text communication presents a promising prospect, as it gives a direct means for individuals to articulate their thoughts and requirements \cite{intro4}. This technology has the potential to revolutionize their lives, enabling them to interact with the world, express themselves creatively, and regain a sense of independence.

Early research in EEG primarily concentrated on areas like emotion recognition and neurological condition studies. For instance, in \cite{literature1}, researchers reviewed how EEG signals could be linked to understanding human emotions by analyzing brain activities. A Similar study \cite{literature2} extended this focus to emotion recognition from EEG data, exploring a range of techniques from initiating emotions through the preprocessing of EEG signals to extracting and classifying features. This research also critically evaluated the strengths and weaknesses of these varied methods. Moving away from emotional analysis, the study \cite{literature3} shifted the focus toward diagnosing Alzheimer’s Disease. It offered an in-depth look into the use of EEG for detecting Alzheimer’s, including a detailed analysis of the complexity found in the EEG signals associated with the disease. However, there's still a gap in the literature: no review articles have yet focused on converting EEG signals to text. This gap is significant, considering the potential applications and advancements this research could bring. 

Considering the significance of EEG-to-text conversion, this review compiles and examines recent research in this burgeoning field. We aim to outline and discuss the techniques used in transforming EEG signals into text, aiming to guide future research directions. The contributions of our review are the following:

\begin{itemize}
    \item Firstly, we address the various challenges faced in decoding EEG signals, providing insight into the complexities inherent in this process.

    \item Secondly, we design a comprehensive taxonomy that categorizes and discusses the range of techniques used in this domain, from initial data collection to the intricacies of model development.

    \item Lastly, we explore potential avenues for future research, identifying gaps in current knowledge and proposing areas that hold promise for further investigation.
\end{itemize}

This review aims not just to go over what's been done but also to spark new ideas and paths in the field of EEG-to-text conversion.

The rest of this paper is structured as follows: In Section II, we delve into the challenges of EEG signal decoding, detailing six distinct types of challenges encountered in this field. Section III presents a taxonomy, concentrating on the techniques employed, from the initial stages of data collection to the intricacies of model development. In Section IV, we highlight potential directions for future research. The paper concludes with Section V, summarizing our findings and final thoughts.

\section{Challenges in EEG Decoding}
\begin{figure*}[t]
    \centering
    \includegraphics[width=1\linewidth]{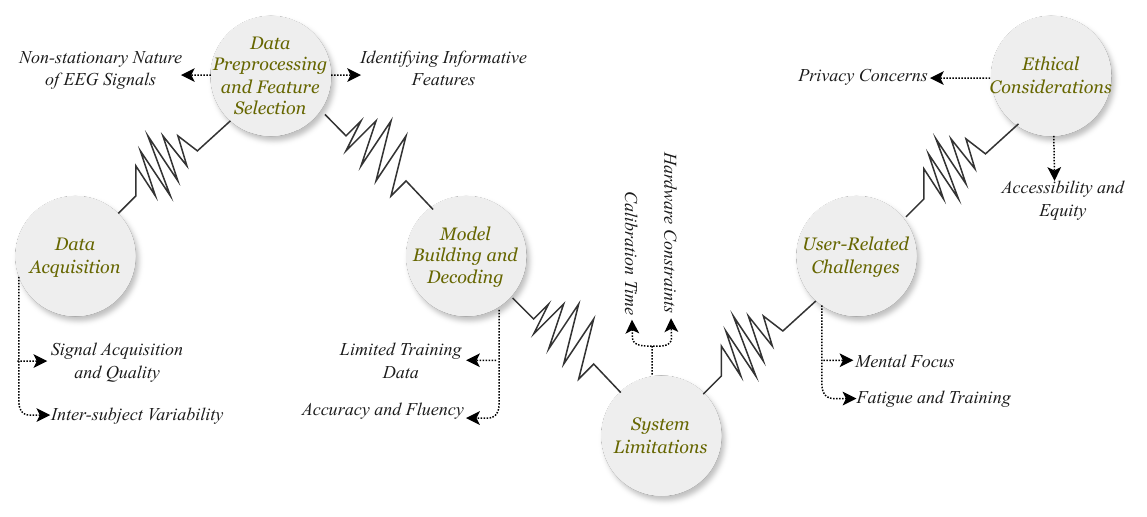}
    \caption{Challenges in EEG Signal Decoding for Text Generation.}
    \vspace{-10pt}
    \label{challenges}
\end{figure*}

Despite the immense potential of EEG decoding for text generation, translating brain activity into written language presents significant challenges. This section delves into the complexities of this process, highlighting the key hurdles researchers face at each stage.  Figure \ref{challenges} presents the challenges of decoding the EEG signals into text. This figure provides a roadmap to understanding the complexities of EEG decoding for text generation. It outlines the various stages involved, from data acquisition to model building, and highlights the key challenges encountered at each step. 

\subsection{Data Acquisition}

\textbf{Signal Acquisition and Quality: }Acquiring clean and high-quality EEG signals is essential for accurate text decoding. However, this process faces several hurdles. The inherent weakness of brain signals compared to background electrical noise, particularly from muscles and power lines, necessitates sophisticated noise reduction techniques \cite{signal_acquisition1,signal_acquisition2}.  Furthermore, significant variability in brain activity patterns between individuals due to anatomical and cognitive differences poses a challenge for developing generalized decoding models that work effectively for everyone \cite{EEG-To-Text}. Even slight head movements can introduce artifacts, further muddying the signal \cite{signal_acquisition1}. Additionally, the limited spatial resolution of EEG, which measures activity from a large brain area, makes it difficult to pinpoint the exact source of language-related activity \cite{signal_acquisition3}. Finally, user comfort and training can be hurdles. Wearing an EEG cap with multiple electrodes can be uncomfortable for some users, and extensive training sessions may be required to learn how to control their brain activity for optimal decoding results \cite{signal_acquisition4}. 

\textbf{Inter-subject Variability: }Despite the potential of EEG for text decoding, a major hurdle lies in inter-subject variability.  Unlike fingerprints, brain activity patterns are highly individualistic. This variation arises from several sources.  Firstly, anatomical differences in brain structure and neuron distribution between people lead to diverse electrical activity patterns \cite{signal_acquisition4}.  Secondly, cognitive styles influence which brain regions are activated during thought processes.  Some individuals may rely heavily on visual processing, while others favor auditory or kinesthetic pathways \cite{EEG-To-Text}.  Finally, even slight variations in how EEG electrodes are placed on the scalp can significantly impact the recorded signals and decoding accuracy across users \cite{signal_acquisition5}.  As a result, decoding models may need to be personalized for each user, as a model trained on one person's EEG data might not perform well for someone with a different brain or cognitive style. 

\subsection{Data Preprocessing and Feature Selection}

\textbf{Non-Stationary Nature of EEG Signals: }Understanding the intricate connection between brain activity and written language is further complicated by the non-stationary nature of EEG signals. Unlike stationary signals with consistent statistical properties over time, EEG signals exhibit dynamic changes \cite{signal_acquisition1}. These variances can arise from internal cognitive changes as users direct their attention towards different elements of the text they intend to produce \cite{non-stationary2}. Additionally, external factors like fatigue or slight head movements might cause temporary fluctuations in the signal \cite{signal_acquisition4}. This non-stationary nature poses challenges for data preprocessing and feature selection. Traditional techniques assuming stationary signals may not effectively capture the time-varying information crucial for accurate text decoding. Researchers are exploring methods like time-frequency analysis and adaptive filtering techniques to account for the non-stationary characteristics of EEG signals and extract the most relevant features for successfully decoding brain activity into text \cite{non-stationary4}.

\textbf{Identifying Informative Features: }In the context of transforming EEG data to text, an important step in data preprocessing and feature selection involves identifying informative features. This process is important because EEG data is characteristically high-dimensional and contains a significant amount of non-informative or redundant information. Efficient feature selection can significantly improve the performance of machine learning models used in this domain. The challenge lies in distinguishing relevant features that are most representative of the underlying cognitive or neural processes from the irrelevant ones. Techniques such as PCA, ICA, and mutual information-based methods have been employed to address this issue \cite{identifying1,identifying2}. These methods aim to reduce the dimensionality of EEG data while retaining the most significant information, facilitating the subsequent machine learning tasks such as classification or regression required for converting EEG signals into text representations. This step is important for applications in BCI, where the goal is to translate neural activity into actionable commands or textual forms \cite{identifying3}.

\subsection{Model Building and Decoding}
\textbf{Limited training data: }The scarcity of training data poses a major obstacle in the creation of efficient EEG-to-text algorithms. Due to the inherent complexity and variability of EEG signals, coupled with the difficulty in collecting large datasets, machine learning models often suffer from inadequate training, leading to poor generalization and performance. This problem is most noticeable in EEG-to-text applications, where the model needs to interpret intricate brain signals and produce accurate textual results. Transfer learning and data augmentation are often used techniques to address this difficulty. Transfer learning involves fine-tuning a pre-trained model on a smaller dataset, while data augmentation artificially enhances the quantity and variability of training datasets \cite{limited1}. Furthermore, the use of generative models to create synthetic EEG data, which resembles real recordings, has shown promise in enhancing the training process (Hartmann et al., 2018). These approaches aim to overcome the limitations imposed by scarce EEG data, thus improving the accuracy and reliability of EEG-to-text conversion models crucial for applications in neural prosthesis and BCI \cite{limited2}.

\textbf{Accuracy and Fluency: }Attaining a high level of accuracy and fluency is still a difficult task when it comes to constructing models and decoding for EEG-to-text conversion. The inherent complexity of interpreting EEG signals, which are often noisy and highly individualized, poses significant difficulties in accurately translating these signals into coherent and fluent text. The accuracy of a model refers to its ability to accurately interpret the EEG signals, while fluency pertains to the naturalness and readability of the generated text. These dual objectives often require sophisticated algorithms that can handle the intricate patterns in EEG data.  Researchers have studied the possibility of deep learning models, specifically recurrent neural networks (RNNs) and attention mechanisms, to improve both accuracy and fluency in this field \cite{accuracy1}. These models can capture temporal dependencies and contextual nuances in the EEG signals, which are crucial for producing precise and fluent text output. However, the trade-off between accuracy and fluency remains a key area of research, as improving one aspect can sometimes be at the expense of the other. This balancing act is critical in applications like real-time communication aids for individuals with speech impairments, where both accuracy and fluency are essential for effective interaction \cite{accuracy2,accuracy3}.

\subsection{System Limitations}
\textbf{Hardware Constraints: } Hardware limitations pose a substantial obstacle in the creation and execution of EEG-to-text systems. The capabilities of the EEG recording equipment directly affect the quality and resolution of EEG signals, which are essential for proper decoding. The availability of high-resolution EEG devices, which offer intricate neurological data, is often restricted due to their high cost and limited accessibility. Consequently, the general utilization of advanced EEG-to-text applications is hindered \cite{Hardware1}. Moreover, these sophisticated devices can be unwieldy and intrusive, impeding their suitability for daily usage. However, portable and user-friendly EEG equipment typically have lower resolution and are more prone to noise and aberrations. These factors can have a negative impact on the performance of EEG-to-text models \cite{Hardware2}. The difficulty is in achieving a harmonious equilibrium between the excellence of EEG data capture and the feasibility and availability of the technology. In addition, the processing and decoding of EEG data in real-time necessitate high-performance processing units, which may not be practical for portable or wearable devices \cite{Hardware3,Hardware4}. The hardware limitation plays a crucial role in assessing the practicality and efficiency of EEG-to-text systems, especially in the context of assistive communication devices designed for individuals with speech or mobility disabilities.

\textbf{Calibration Time: } Calibration time poses a substantial challenge in developing and implementing EEG-to-text models. Calibrating EEG devices to individual users is essential for accurately decoding neural signals. However, this process can be time-consuming and demands substantial human exertion. The brainwave patterns of each individual are distinct, and the EEG system must be precisely calibrated to these patterns to achieve successful communication or control. The calibration process often requires the user to engage in specific cognitive tasks regularly, enabling the system to learn and adjust to their EEG signals \cite{calibration1, Hardware4}. The calibration process for EEG-to-text systems might be challenging in terms of time and effort, especially when considering their applicability to individuals with disabilities or in time-critical scenarios. Efforts to decrease the time required for calibration while also upholding or enhancing the precision and dependability of the system are a crucial focus of research. Researchers are currently investigating emerging methods, such as adaptive algorithms and transfer learning, to tackle this difficulty. These methods involve using calibration data from one user to help calibrate the system for another user \cite{calibration3,Hardware3}. The objective of these techniques is to enhance the user-friendliness and accessibility of EEG-to-text systems, hence expanding their range of potential applications.

\subsection{User Related Challenges}

\textbf{Mental Focus: }The use of EEG-to-text systems faces a notable user-related barrier in terms of mental attention. The success of these systems heavily depends on the user's capacity to sustain unwavering mental concentration, as fluctuations in attention can result in significant deviations in EEG signal patterns, thus impacting the precision of signal interpretation and text generation. This is particularly challenging because maintaining a high concentration level over extended periods is difficult for most individuals. Fatigue, distraction, and cognitive load can all adversely impact the user's ability to generate stable and clear EEG signals \cite{mental1}. Moreover, the requirement for sustained mental focus can be especially demanding for users with cognitive or attention impairments, limiting the accessibility and usability of EEG-to-text technologies for these populations. Researchers are exploring various strategies to address this challenge, such as developing more robust algorithms that can cope with fluctuating levels of user attention and integrating adaptive learning systems that adjust to the user's mental state \cite{Hardware4}. The purpose of these developments is to improve the robustness of EEG-to-text systems against fluctuations in mental concentration, thus increasing their practicality and efficacy for a wider spectrum of users.

\textbf{Fatigue and Training: }User-related issues in EEG-to-text communication systems include fatigue and the requirement for substantial training. Users often experience fatigue during prolonged use of EEG systems, as the process of generating consistent and accurate neural signals for text communication requires considerable mental effort and concentration.  Fatigue can result in a deterioration of the quality of the EEG signals, which in turn affects the function of the system \cite{fatigue1}. Furthermore, the requirement for extensive training to use EEG-to-text systems efficiently can be a barrier to their widespread adoption, particularly for individuals with disabilities or those who lack the time and resources for lengthy training sessions. The training process involves users learning to generate distinct neural patterns that the system can recognize and translate into text, which can be a time-consuming and demanding task \cite{fatigue3}. To address these challenges, researchers are focusing on developing more intuitive and user-friendly interfaces, as well as adaptive algorithms that require less user training and are more resilient to the effects of fatigue \cite{Hardware4}. 

\subsection{Ethical Considerations}

\textbf{Privacy Concerns: } Privacy concerns constitute a critical ethical challenge in the realm of EEG-to-text technology. EEG data, which might disclose private and confidential information about a person's mental condition, thoughts, or intentions, presents substantial concerns regarding privacy \cite{ethical1}. Preserving the confidentiality and security of this data is of utmost importance, as breaches could result in unauthorized access and exploitation of personal information. This is particularly relevant in the context of BCIs, where EEG data is used for direct communication or control. The risk of eavesdropping or hacking into these systems poses a serious threat to user privacy \cite{ethical2}. To address these concerns, researchers and developers are exploring various data protection strategies, such as advanced encryption methods and strict access controls, to safeguard against unauthorized data access and ensure compliance with privacy regulations \cite{ethical3}. Furthermore, ethical guidelines and frameworks are being developed to guide the responsible use of EEG data and protect individuals' privacy rights in the use of EEG-to-text and other BCI technologies \cite{ethical4}.

\textbf{Accessibility and Equity: }Ensuring accessibility and equity are crucial ethical considerations while developing and using EEG-to-text technology. The potential of these systems to provide communication aids for individuals with disabilities highlights the need for inclusive design and equitable access. Nevertheless, there is a potential danger that these technologies may exacerbate the disparity between individuals who have access to sophisticated medical and assistive technologies and those who do not, owing to factors such as expense, technological proficiency, and the presence of healthcare facilities \cite{accessibility1}. Furthermore, the development of EEG-to-text systems frequently prioritizes the preferences of a typical user, possibly neglecting the distinct demands of individuals with different abilities and backgrounds. This lack of inclusivity can result in technologies that are not equally accessible or beneficial to all potential users \cite{accessibility2}. Addressing these issues, it is essential to focus on developing EEG-to-text and other BCI technologies using universal design principles. This involves considering the varied requirements and situations of different user groups. Furthermore, policy measures and funding initiatives could play a significant role in ensuring equitable access to these technologies, particularly for underserved and marginalized communities \cite{accessibility3}.

\section{EEG to Generative Model Pipeline}
\label{sec:8}

\begin{figure*}[t]
    \centering
    \includegraphics[width=1\linewidth]{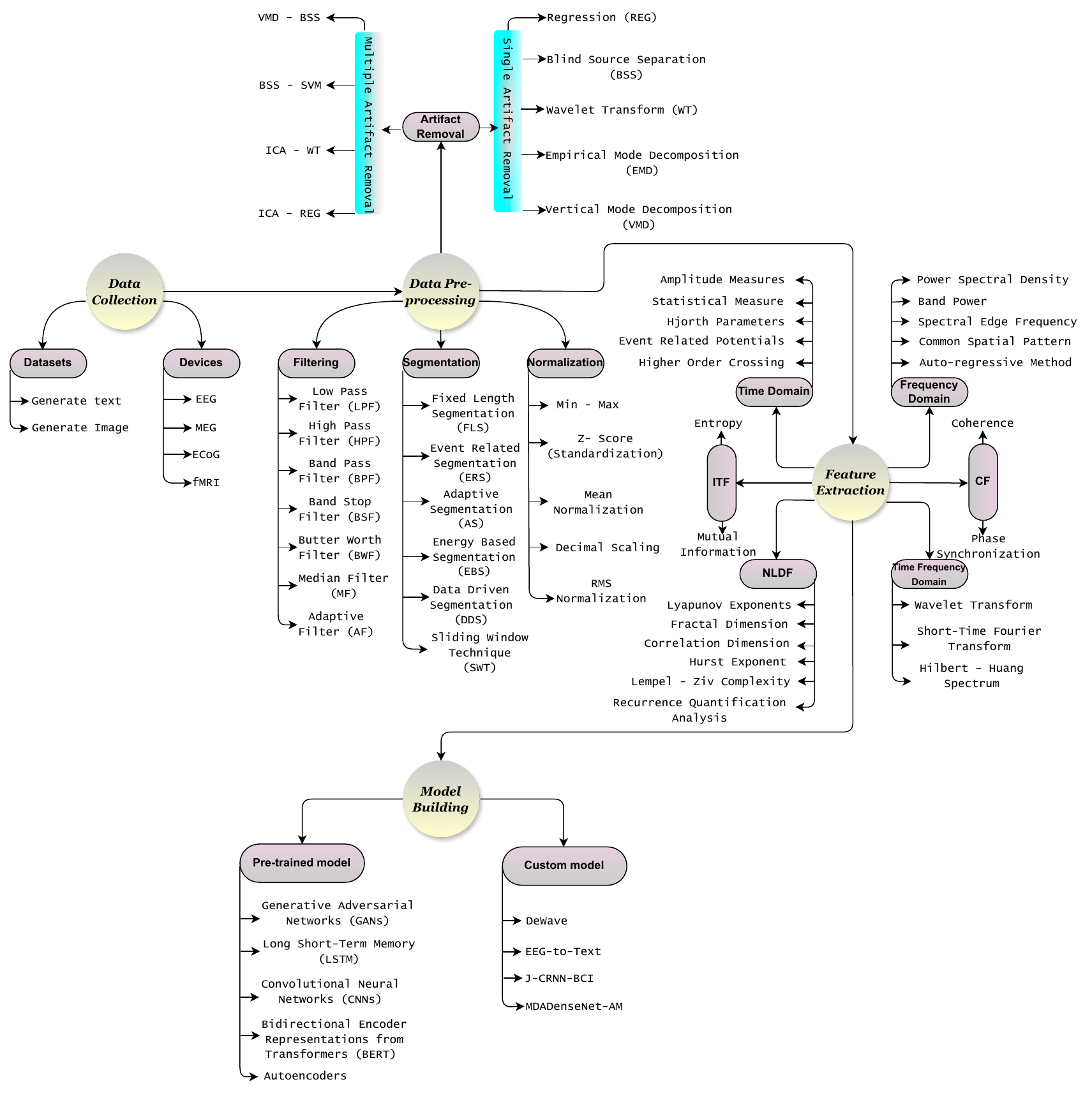}
    \caption{Taxonomy of EEG Signal Processing for Text and Image Generation.}
    \vspace{-10pt}
    \label{taxonomy}
\end{figure*}

Extracting meaningful information from electroencephalogram (EEG) signals is crucial for various applications in neuroscience, brain-computer interfaces, and clinical diagnosis. This process typically involves several stages, as illustrated in Figure \ref{taxonomy}. The initial stage focuses on data acquisition, where electrodes record EEG signals from the scalp. Subsequently, preprocessing techniques are employed to remove noise and artifacts, ensuring the quality of the data for further analysis. After the EEG signals have been preprocessed, feature extraction is a key step in turning them into useful features that capture the right characteristics for the application. This article delves into various feature extraction techniques commonly employed in EEG signal processing. 

\subsection{Data Acquisition}

The data acquisition process encompasses two key components: the dataset and the device. The dataset aspect involves an examination of existing data previously utilized for converting brain signals into text and images. This part of the process focuses on how this data was initially collected and processed to facilitate the transformation of neural activity into comprehensible formats like text and visual imagery. Conversely, the device segment delves into the various instruments employed to gather data directly from the brain. This includes an array of technologies ranging from non-invasive tools like EEG, Magnetoencephalography (MEG), and  Functional MRI (fMRI) to more invasive methods such as Electrocorticography (ECoG).

\subsubsection{Datasets}

\paragraph{Generate Text }Many researchers have made significant contributions to the field of brain signal decoding into text, resulting in the publication of various datasets. These datasets have been gathered using both invasive techniques, like ECoG, and non-invasive techniques, including EEG and fMRI. Notably, ZuCo 1.0 \cite{dataset1} and ZuCo 2.0 \cite{dataset2} are prominent EEG-based datasets collected using setups with 128 channels. In contrast, when utilizing fMRI for data collection \cite{dataset3,dataset4}, researchers often employ 32-channel head coils. Additionally, ECoG, a widely used method for brain signal acquisition \cite{dataset5,dataset6}, typically involves the use of 16 channels during experiments. Table \ref{gen_text_table} lists the datasets utilized for decoding human brain signals into text. 

\begin{table*}[]
\centering
\caption{Description of Publicly Available Datasets Used for Text Generation}
\label{gen_text_table}
\begin{tabular}{|c|cccc|c|c|}
\hline
\multirow{2}{*}{\textbf{Datasets}} & \multicolumn{4}{c|}{\textbf{Participants}}                                                                                                                         & \multirow{2}{*}{\textbf{Channels}}                                                       & \multirow{2}{*}{\textbf{Stimulus}} \\ \cline{2-5}
                                   & \multicolumn{1}{c|}{\textit{\textbf{Male}}} & \multicolumn{1}{c|}{\textit{\textbf{Female}}} & \multicolumn{1}{c|}{\textit{\textbf{Total}}} & \textit{\textbf{Age}} &                                                                                          &                                    \\ \hline
ZuCo 1.0 \cite{dataset1}                          & \multicolumn{1}{c|}{7}                      & \multicolumn{1}{c|}{5}                        & \multicolumn{1}{c|}{12}                      & 22-54                 & EEG (128)                                                                                      & 1107 English sentences             \\ \hline
ZuCo 2.0 \cite{dataset2}                           & \multicolumn{1}{c|}{9}                      & \multicolumn{1}{c|}{10}                       & \multicolumn{1}{c|}{19}                      & 23-54                 &  EEG (128)                                                                                      & 739 English sentences              \\ \hline
fMRI Image \cite{dataset3}                         & \multicolumn{1}{c|}{-}                      & \multicolumn{1}{c|}{-}                        & \multicolumn{1}{c|}{5}                       & 21-50                 & 32-channel head coil                                                                     & 15 subjects with 540 scans         \\ \hline
fMRI Image  \cite{dataset4}                       & \multicolumn{1}{c|}{-}                      & \multicolumn{1}{c|}{-}                        & \multicolumn{1}{c|}{-}                       & -                     & fMRI                                                                                     & 180 fMRI Images                    \\ \hline
ECoG Data     \cite{dataset5}                     & \multicolumn{1}{c|}{-}                      & \multicolumn{1}{c|}{-}                        & \multicolumn{1}{c|}{7}                       & -                     & \begin{tabular}[c]{@{}c@{}}Eight16-channel g.USBamp \\ biosignal amplifiers\end{tabular} & 4381 second voice recording        \\ \hline
ECoG Data   \cite{dataset6}                       & \multicolumn{1}{c|}{1}                      & \multicolumn{1}{c|}{4}                        & \multicolumn{1}{c|}{5}                       & 29-49                 & 16 Channel                                                                               & 4053 English sentences             \\ \hline
\end{tabular}
\end{table*}

\paragraph{Generate Image }Many other researchers have focused on creating images from brain signals, primarily utilizing non-invasive methods such as EEG \cite{image1,image2,image3} and fMRI \cite{image1,image4,image5,image6}. In these studies, the stimuli primarily consist of images, though a few studies have also explored the use of text as stimuli. This exploration into brain signal-based image generation is a growing field, delving into the complex relationship between neural activity and visual representation. By analyzing brain responses to visual stimuli, these studies aim to reconstruct or generate images that correlate with the observed brain activity, providing insights into how the brain processes and interprets visual information. Table \ref{gen_image_table} provides a comprehensive list of datasets used to convert human brain signals into images.

\begin{table*}[]
\centering
\caption{Description of Publicly Available Datasets Used for Image Generation}
\label{gen_image_table}
\begin{tabular}{|c|cccc|c|c|}
\hline
\multirow{2}{*}{\textbf{Datasets}} & \multicolumn{4}{c|}{\textbf{Participants}}                                                                                                                         & \multirow{2}{*}{\textbf{Channels}}                                                           & \multirow{2}{*}{\textbf{Stimulus}} \\ \cline{2-5}
                                   & \multicolumn{1}{c|}{\textit{\textbf{Male}}} & \multicolumn{1}{c|}{\textit{\textbf{Female}}} & \multicolumn{1}{c|}{\textit{\textbf{Total}}} & \textit{\textbf{Age}} &                                                                                              &                                    \\ \hline
EEG + fMRI    \cite{image1}                     & \multicolumn{1}{c|}{-}                      & \multicolumn{1}{c|}{-}                        & \multicolumn{1}{c|}{5}                       & 25-27                 & \begin{tabular}[c]{@{}c@{}}EEG(64), 3.0-Tesla Siemens \\ MAGNETOM Verio scanner\end{tabular} & 50 images in 20 category           \\ \hline
EEG data     \cite{image2}                      & \multicolumn{1}{c|}{-}                      & \multicolumn{1}{c|}{-}                        & \multicolumn{1}{c|}{23}                      & 15-40                 & EEG (14)                                                                                     & 20 text and10 non-text items       \\ \hline
EEG data    \cite{image3}                       & \multicolumn{1}{c|}{5}                      & \multicolumn{1}{c|}{1}                        & \multicolumn{1}{c|}{6}                       & -                     & EEG (128)                                                                                    & 2000 image                         \\ \hline
fMRI      \cite{image4}                         & \multicolumn{1}{c|}{2}                      & \multicolumn{1}{c|}{1}                        & \multicolumn{1}{c|}{3}                       & 23-33                 & \begin{tabular}[c]{@{}c@{}}3.0-Tesla Siemens MAGNETOM\\  Verio scanner\end{tabular}          & 1200 natural images                \\ \hline
fMRI     \cite{image5}                          & \multicolumn{1}{c|}{-}                      & \multicolumn{1}{c|}{-}                        & \multicolumn{1}{c|}{2}                       & -                     & 4T INOVAMR scanner                                                                           & 1870 images                        \\ \hline
NSD  \cite{image6}                              & \multicolumn{1}{c|}{-}                      & \multicolumn{1}{c|}{-}                        & \multicolumn{1}{c|}{8}                       & -                     & fMRI                                                                                         & 27750 fMRI-image                   \\ \hline
\end{tabular}
\end{table*}

\subsubsection{Devices }Various companies offer a variety of devices for monitoring human brain signals, utilizing both invasive and non-invasive methods to collect data. In non-invasive techniques, electrodes are positioned on the scalp without surgical intervention. This approach is commonly used due to its safety and ease of application. In contrast, invasive techniques involve a surgical process to place electrodes directly on or within the brain, providing more direct and often more detailed brain signal readings. Table \ref{device_name} provides a comprehensive overview of the most popular devices currently used. This includes various organizations such as NeuroScan \cite{NeuroScan}, Brain Products \cite{brain}, BioSemi \cite{biosemi}, Emotiv \cite{EMOTIV}, NeuroSky \cite{NeuroSky}, ANT Neuro \cite{ANT}, ABM \cite{ABM}, and OpenBCI \cite{OpenBCI}. Each organization designs devices with varying ranges of electrical channels and frequencies.

\begin{table}[h]
\begin{scriptsize}
\centering
\caption{Name of Companies Offering Brain Signal Acquisition Devices with Varied Functionalities}
\label{device_name}
\begin{tabular}[c]{|p{2cm}|p{1.7cm}|p{1.2cm}|p{.2cm}p{.2cm}p{.4cm}|} 
\hline
\multirow{2}{*}{\textbf{Name}} & \multirow{2}{*}{\textit{\textbf{\begin{tabular}[c]{@{}c@{}}Sampling \\ Rate\end{tabular}}}} & \multirow{2}{*}{\textit{\textbf{Channel}}} & \multicolumn{3}{c|}{\textit{\textbf{Devices}}}                                                                   \\ \cmidrule(l){4-6} 
                               &                                                                                             &                                            & \multicolumn{1}{c|}{\textit{\textbf{EEG}}} & \multicolumn{1}{c|}{\textit{\textbf{MEG}}} & \textit{\textbf{fMRI}} \\ \hline
NeuroScan   \cite{NeuroScan}                   & 500 Hz                                                                                      & 32, 64, 128, 256                           & \multicolumn{1}{c|}{\cmark} & \multicolumn{1}{c|}{\cmark} & \cmark  \\ \hline
Brain Products     \cite{brain}            & \begin{tabular}[c]{@{}c@{}}1000 Hz, 500 Hz, \\ 250 HZ\end{tabular}                          & 8, 16, 32, 64                              & \multicolumn{1}{c|}{\cmark} & \multicolumn{1}{c|}{\xmark} & \cmark  \\ \hline
BioSemi       \cite{biosemi}                 & 2048 Hz, 1024 Hz                                                                            & 32, 64, 128, 256                           & \multicolumn{1}{c|}{\cmark} & \multicolumn{1}{c|}{\xmark} & \xmark  \\ \hline
Emotiv           \cite{EMOTIV}              & 2048 Hz, 128 Hz                                                                             & 2, 5, 14, 32                               & \multicolumn{1}{c|}{\cmark} & \multicolumn{1}{c|}{\xmark} & \xmark  \\ \hline
NeuroSky     \cite{NeuroSky}                  & 512 Hz                                                                                      & 12                                         & \multicolumn{1}{c|}{\cmark} & \multicolumn{1}{c|}{\xmark} & \xmark  \\ \hline
ANT Neuro     \cite{ANT}                 & 16 kHz, 2048 Hz                                                                             & 32, 64, 128, 256                           & \multicolumn{1}{c|}{\cmark} & \multicolumn{1}{c|}{\cmark} & \cmark  \\ \hline
ABM       \cite{ABM}                     & 4 kHz, 2048 Hz                                                                              & 24, 32, 64                                 & \multicolumn{1}{c|}{\cmark} & \multicolumn{1}{c|}{\xmark} & \xmark  \\ \hline
OpenBCI      \cite{OpenBCI}                  & \begin{tabular}[c]{@{}c@{}}125 Hz, 200 Hz, \\ 250 Hz\end{tabular}                           & 4, 8, 16                                   & \multicolumn{1}{c|}{\cmark} & \multicolumn{1}{c|}{\xmark} & \xmark  \\ \hline
\end{tabular}
\end{scriptsize}
\end{table}

\subsection{Data Pre-processing}

\subsubsection{Artifact Removal}
During EEG signal collection, artifacts are a common issue. These artifacts can be of two types: psychological and non-psychological. Non-psychological artifacts originate from external sources such as electrode malfunctions, the movement of cables, or poor connections in channels. On the other hand, physiological artifacts are caused by internal electrical signals within the body.  It is necessary to remove these unwanted signals. Some physiological artifacts, like those caused by skin and sweat, can be eliminated during neural activity recording. For example, wearing a cooling ventilation vest helps regulate body temperature during physical activities. Power line artifacts can be effectively filtered out using notch filtering techniques. Moreover, movement-related distortions in the EEG signal, caused by head or body motion and the resultant electrode and cable movement, can be reduced by employing a double-layer cap.

In the field of EEG signal processing, various strategies have been developed to eliminate artifacts, as depicted in Figure 1. These techniques are broadly categorized into two groups: single artifact removal and multiple artifact removal methods. Common single artifact removal methods utilized by researchers include techniques such as Regression (REG) \cite{REG}, Blind Source Separation (BSS) \cite{BSS}, Wavelet Transform (WT) \cite{WT1,WT2}, Empirical Mode Decomposition (EMD) \cite{EMD}, and Vertical Mode Decomposition (VMD) \cite{VMD}. In recent times, there has been a growing trend among researchers to combine two single artifact removal methods to enhance the efficacy of artifact removal from EEG signals. This approach has led to the development of sophisticated hybrid techniques such as VMD-BSS \cite{VMD-BSS}, BSS-SVM \cite{BSS-SVM}, ICA-WT \cite{ICA-WT}, and REG-ICA \cite{REG-ICA}. These multiple artifact removal techniques offer a more robust and comprehensive approach, addressing a broader range of artifacts compared to single artifact removal methods. Consequently, they have gained popularity in the research community for their improved ability to clean EEG signals from various types of unwanted interference. Table \ref{artifact_table} delineates a range of artifact mitigation strategies employed in EEG signal analysis as explored by various researchers.

\begin{table*}[]
\begin{scriptsize}
\caption{Comparative Overview of Artifact Removal Techniques in EEG Signal Processing}
\label{artifact_table}
\begin{tabular}{|c|c|c|c|c|c|c|}
\hline
\textbf{Ref.} & \textbf{Type of Artifact}                                                                                                              & \textbf{ \begin{tabular}[c]{@{}c@{}} Artifact Removal\\ Technique\end{tabular}} & \textbf{\begin{tabular}[c]{@{}c@{}}Single Artifact\\ Removal\end{tabular}} & \textbf{\begin{tabular}[c]{@{}c@{}}Multiple Artifact\\ Removal\end{tabular}} & \textbf{Application Area}                                                                                                & \textbf{Performance Metrics}                                                                                                              \\ \hline
                    \cite{WT1}
                   & \begin{tabular}[c]{@{}c@{}}Motion, Eye blinking, EMG\\ Artifact\end{tabular}                                                           & WT                                  & \cmark                                                      & \xmark                                                        & Clinical Monitoring                                                                                                      & Signal-to-noise ratio (SNR)                                                                                                               \\ \hline
                   \cite{WT2}
                   & \begin{tabular}[c]{@{}c@{}}Eye movement and blinking, \\ Swallowing, Chewing, \\ Limb movement, Body\\  movement artifact\end{tabular} & WT                                  & \cmark                                                      & \xmark                                                        & Brain-Computer Interface (BCI)                                                                                           & \begin{tabular}[c]{@{}c@{}}Signal-to-noise ratio (SNR),\\  root mean square error (RMSE),\\ Lambda\end{tabular}                           \\ \hline
                   \cite{BSS}
                   & \begin{tabular}[c]{@{}c@{}}Ocular, cardiac, muscle\\  and powerline artifacts\end{tabular}                                             & BSS                                 & \cmark                                                      & \xmark                                                        & \begin{tabular}[c]{@{}c@{}}Epileptic spike and seizure \\ detection and brain-computer\\  interfaces (BCIs)\end{tabular} & Signal-to-artifact ratio (SAR)                                                                                                            \\ \hline
                   \cite{BSS-SVM}
                   & \begin{tabular}[c]{@{}c@{}}Eye blinks and heart \\ rhythm artifacts\end{tabular}                                                       & BSS - SVM                           & \xmark                                                      & \cmark                                                        & \begin{tabular}[c]{@{}c@{}}Neurology and brain\\  research\end{tabular}                                                  & Signal-to-Artifact Ratio (SAR)                                                                                                            \\ \hline
                   \cite{ICA-WT}
                   & Electrocardiographic                                                                                                                   & ICA - WT                            & \xmark                                                      & \cmark                                                        & Biomedical engineering                                                                                                   & Signal-to-Artifact Ratio (SAR)                                                                                                            \\ \hline
                   \cite{REG-ICA}
                   & \begin{tabular}[c]{@{}c@{}}Eye movements and \\ blink artifacts\end{tabular}                                                           & REG - ICA                           & \xmark                                                      & \cmark                                                        & Biomedical signal processing                                                                                             & \begin{tabular}[c]{@{}c@{}}Artifact to signal ratio (ASR),\\  Root mean square error (RMSE), \\ Power spectral density (PSD)\end{tabular} \\ \hline
                   \cite{REG}
                   & Eye blinking artifacts                                                                                                                 & REG                                 & \cmark                                                      & \xmark                                                        & Clinical Neurology                                                                                                       & \begin{tabular}[c]{@{}c@{}}Root Mean Square Error (RMSE),\\ Mutual Information (MI)\end{tabular}                                          \\ \hline
                   \cite{EMD}
                   & Muscle artifacts                                                                                                                       & EMD                                 & \cmark                                                      & \xmark                                                        & \begin{tabular}[c]{@{}c@{}}Patients with movement \\ disorders\end{tabular}                                              & \begin{tabular}[c]{@{}c@{}}Relative Root Mean Square\\  Error (RRMSE)\end{tabular}                                                        \\ \hline
                   \cite{VMD}
                   & \begin{tabular}[c]{@{}c@{}}Eye blinking, flutters and \\ lateral eye movements \\ artifacts\end{tabular}                               & VMD                                 & \cmark                                                      & \xmark                                                        & \begin{tabular}[c]{@{}c@{}}Neurophysiology and clinical \\ neurology\end{tabular}                                        & \begin{tabular}[c]{@{}c@{}}Multiscale modified sample \\ entropy (mMSE)\end{tabular}                                                      \\ \hline
                   \cite{VMD-BSS}
                   & \begin{tabular}[c]{@{}c@{}}Muscular activity, \\ heartbeat, and eye \\ movements\end{tabular}                                          & VMD - BSS                           & \xmark                                                      & \cmark                                                        & N/A                                                                                                                      & \begin{tabular}[c]{@{}c@{}}Euclidean Distance (ED), \\ Spearman Correlation \\ Coefficient (SCC)\end{tabular}                             \\ \hline
\end{tabular}
\end{scriptsize}
\end{table*}

\subsubsection{Filtering } Filtering is another important step of EEG signal processing that is used to enhance the quality of the signal by minimizing noise and interference that can obscure the actual signals. There are many ways from which the unwanted signals come, such as power line noise, environmental electromagnetic interference, and physiological artifacts like muscle movements or eye blinks. By removing these unwanted signals, filtering makes it possible to separate the meaningful brainwave patterns that are useful in research. This isolation is particularly important because EEG signals are typically weak and can be easily contaminated by extraneous noise.

EEG signals can be categorized into two primary types: linear and nonlinear.  Linear filters, such as low-pass \cite{low-pass}, high-pass \cite{high-pass}, band-pass \cite{band-pass}, band-stop \cite{band-stop}, and butter-worth filters \cite{butter-worth}, are commonly used to remove frequency components outside of the desired range. For instance, low-pass filters are employed to filter out high-frequency disturbances, while high-pass filters are designed to eliminate low-frequency noise. Conversely, bandpass filters are adept at removing noise that lies outside a designated frequency band. Nonlinear filters, on the other hand, include adaptive filters \cite{adaptive} and median filters \cite{median}, which are more complex and can be tailored to the specific characteristics of the EEG signal. Adaptive filters are particularly useful in scenarios where the signal or noise characteristics are changing over time, as they can dynamically adjust their filtering parameters. The choice of filter type and settings mainly depends on the characteristics of the EEG data and the goals of the analysis. This flexibility and specificity in filtering ensure that the most relevant and accurate information is extracted from the EEG data, significantly enhancing the quality of the signals.

\subsubsection{Segmentation} Segmentation in EEG signal processing is an important preprocessing step that involves dividing continuous EEG data into smaller, more manageable segments. It's important for many different reasons. Firstly, it enhances noise reduction, as segmenting the signal allows for easier identification and elimination of artifacts and interference \cite{adaptive}. Secondly, it facilitates event-related analysis, particularly in cognitive or sensory studies, by enabling precise examination of brain responses to specific stimuli \cite{segmentation1}. Moreover, considering the non-stationary nature of EEG signals, segmentation helps analyze data within shorter, more statistically uniform intervals, ensuring more accurate interpretations. Finally, from a practical standpoint, segmentation increases computational efficiency by breaking down lengthy EEG recordings into smaller, more computationally manageable units. This preprocessing step is particularly important in studies where temporal precision and data quality are paramount, such as in cognitive neuroscience research, clinical diagnostics, and real-time BCI systems.

Researchers employ various segmentation techniques to mitigate noise in EEG signal processing, each with its own unique approach and application. Fixed-Length Segmentation (FLS) \cite{FLS} is widely used for its simplicity, segmenting the EEG data into equal, predetermined lengths, thus providing a uniform framework for analysis. Event-related Segmentation (ERS) \cite{ERS} is tailored for cognitive and sensory studies, where it segments the signal based on specific stimuli or events. Adaptive Segmentation (AS) \cite{adaptive} offers flexibility by adjusting the segment length in response to the signal's characteristics, making it ideal for non-stationary data. Energy-Based Segmentation (EBS) \cite{EBS} relies on the signal's energy content to determine segment boundaries, which is particularly useful in detecting and analyzing high-energy neural events. Data-Driven Segmentation (DDS) \cite{DDS} employs algorithms to segment the data based on inherent features of the EEG signal itself, thus adapting to the signal's natural structure. Finally, the Sliding Window Transform (SWT) \cite{SWT} is a complex method that uses wavelet transforms to deal with non-stationary data. It offers a multi-resolution analysis that is useful for pulling out complex EEG signals' nuanced features. Each of these techniques offers distinct advantages and is chosen based on the specific requirements of the study.

\subsubsection{Normalization } Normalization is another important step in EEG signal processing, mainly due to the high variability of EEG signals both within and between individuals. Normalization helps to mitigate inherent noise in EEG data, such as electrical interference or artifacts from muscle movements, by scaling the EEG signals. EEG signals ensure that outliers or variations in amplitude do not dominate the signal; normalization facilitates the extraction of meaningful biomarkers from the EEG data,  which are vital for subsequent analysis and interpretation.

Researchers employ various techniques to normalize EEG signals, each with its own distinct approach to standardizing the data. Min-Max \cite{min-max} normalization rescales the data to a fixed range, typically [0, 1], ensuring that each signal falls within a standardized band. Z-score \cite{z-score} normalization, also known as standard score normalization, centers the data around the mean with a unit standard deviation, thereby addressing scale and distribution shape issues. Mean normalization \cite{mean-norm} adjusts the data values to revolve around the mean, effectively balancing the dataset around a central value. Decimal normalization \cite{decimal-scaling} shifts the decimal point of values, standardizing them based on their magnitude, which is particularly useful for varying signal amplitudes. Lastly, RMS (Root Mean Square) normalization \cite{rms-norm} scales the signal by the square root of the mean squared value, often used to maintain the signal's energy across different conditions. These normalization methods collectively enhance EEG data's comparability and consistency, facilitating more accurate analyses.

\begin{table}[]
\centering
\caption{Comparison of Feature Extraction Methods for EEG Signal Processing.}
\label{feature_extraction}
\begin{tabular}{|c|c|c|c|c|c|c|}
\hline
Ref. & \begin{tabular}[c]{@{}c@{}}Time \\ Domain\end{tabular} & \begin{tabular}[c]{@{}c@{}}Frequency \\ Domain\end{tabular} & \begin{tabular}[c]{@{}c@{}}Time Frequency \\ Domain\end{tabular} & NLDF                  & ITF                   & \begin{tabular}[c]{@{}c@{}}CF\end{tabular} \\ \hline
\cite{Tim_Freq_TF}    & \cmark                                  & \cmark                                      & \cmark                                            & \xmark & \xmark & \xmark                                            \\ \hline
\cite{TF}    & \xmark                                  & \xmark                                      & \cmark                                            & \xmark & \xmark & \xmark                                            \\ \hline
\cite{Freq_TF}    & \xmark                                  & \cmark                                      & \cmark                                            & \xmark & \xmark & \xmark                                            \\ \hline
\cite{Freq}    & \xmark                                  & \cmark                                      & \xmark                                            & \xmark & \xmark & \xmark                                            \\ \hline
\cite{coh_CF}    & \xmark                                  & \xmark                                      & \xmark                                            & \xmark & \xmark & \cmark                                            \\ \hline
\cite{Entr_ITF}    & \xmark                                  & \xmark                                      & \xmark                                            & \xmark & \cmark & \xmark                                            \\ \hline
\cite{PS_CF}    & \xmark                                  & \xmark                                      & \xmark                                            & \xmark & \xmark & \cmark                                            \\ \hline
\cite{FD_NLDF}    & \xmark                                  & \xmark                                      & \xmark                                            & \cmark & \xmark & \xmark                                            \\ \hline
\cite{AM_Time}    & \cmark                                  & \xmark                                      & \xmark                                            & \xmark & \xmark & \xmark                                            \\ \hline
\cite{PSD_Freq}    & \xmark                                  & \cmark                                      & \xmark                                            & \xmark & \xmark & \xmark                                            \\ \hline
\cite{HHS_TFD}    & \xmark                                  & \xmark                                      & \cmark                                            & \xmark & \xmark & \xmark                                            \\ \hline
\end{tabular}
\end{table}

\subsection{Feature Extraction }
Feature extraction is a critical step in EEG signal processing due to the inherent complexity and high dimensionality of the data. It acts as a bridge between raw EEG recordings and meaningful insights. Extracting relevant and informative features from various domains (time, frequency, time-frequency, and space) compresses the data while preserving key information crucial for subsequent analysis and interpretation. This allows researchers to effectively utilize ML algorithms and unlock the hidden potential of EEG data for various applications in neuroscience, BCI, and clinical diagnosis. Figure \ref{feature_extraction} shows a comparison of different feature extraction methods that were used in previous research. 

\subsubsection{Time Domain } Time-domain analysis is a fundamental approach for extracting features from EEG signals. It focuses on characterizing the signal's behavior directly over time, offering insights into its amplitude, variability, and rhythmicity. This domain proves valuable for capturing transient events and quantifying specific characteristics within the signal. Common time domain techniques include Amplitude measures \cite{AM_Time}, Statistical measures \cite{SM_Time}, Hjorth parameters \cite{HP_Time, HP_HOC_Time}, Event-related potentials (ERPs) \cite{ERP_Time}, and Higher-order crossing (HOC) \cite{HP_HOC_Time}.  Each method contributes uniquely to the comprehensive analysis of EEG signals, making them indispensable in extracting meaningful information from complex brain activity. 

\subsubsection{Frequency Domain } The frequency domain approach is another highly effective technique for feature extraction in EEG signal processing. This method transforms EEG signals to analyze their frequency components, providing a different perspective compared to time-domain analysis. This domain proves valuable for understanding the dominant rhythms associated with various brain activities and identifying event-related changes in the frequency spectrum. Key methods utilized in this domain include Power Spectral Density \cite{PSD_Freq}, Band Power \cite{BP_Freq}, Spectral Edge Frequency \cite{SEF_Freq}, Common Spatial Pattern \cite{CSP_Freq}, and the Auto-regressive Method \cite{ARM_Freq}. Each of these plays a significant role in analyzing and interpreting the complex frequency-based characteristics of EEG data.

\subsubsection{Time Frequency Domain } While both time-domain and frequency-domain analyses offer valuable insights into EEG signals, they each provide a limited perspective. Time-domain analysis excels at capturing temporal dynamics but lacks resolution in the frequency domain. Conversely, frequency-domain analysis excels at revealing spectral characteristics but fails to capture how these characteristics change over time. 

To overcome these limitations, researchers often employ time-frequency domain analysis. This approach provides a comprehensive understanding of EEG signals by simultaneously analyzing their temporal and spectral information. By decomposing the signal into its time-frequency components, researchers can gain insights into how the frequency content of the signal evolves over time. Among the most prominent techniques in this domain are the Wavelet Transform \cite{WT1, WT2}, Short-Time Fourier Transform (STFT) \cite{STFT_TFD}, and Hilbert-Huang Spectrum \cite{HHS_TFD}. 

\subsubsection{Non-linear Dynamic Features (NLDF) } NLDF is a new way to look at EEG signals beyond the usual time-domain, frequency-domain, and time-frequency methods. This method explores the inherent non-linearity and complexity of brain activity, which linear models are unable to capture fully. NLDF techniques focus on extracting features that characterize the dynamic behavior of the EEG signal over time. These features often involve measures of complexity, chaos, and synchronization, providing insights into the underlying mechanisms of brain function. NLDF such as Lyapunov Exponents \cite{LE_NLDF}, Fractal Dimension \cite{FD_NLDF}, Correlation Dimension \cite{CD_HE_NLDF}, Hurst Exponent \cite{CD_HE_NLDF}, Lempel-Ziv Complexity \cite{LZC_NLDF}, and Recurrence Quantification Analysis \cite{RQA_NLDF} are pivotal in EEG signal processing for capturing the complex, dynamic behavior of the brain's electrical activity.

\subsubsection{Information Theoretic Features (ITF) } In recent years, ITF has emerged as a powerful tool for extracting meaningful information from EEG signals. Entropy \cite{Entr_ITF}, a fundamental ITF measure, evaluates the unpredictability or randomness of the signal, offering insights into the complexity of neural dynamics. It is particularly useful in assessing the regularity and predictability of EEG signals, which can be pivotal in differentiating between different neurological states or conditions. On the other hand, Mutual Information \cite{MI_ITF} measures the amount of information shared between two signals, reflecting the degree of statistical dependency and potential interaction between different brain regions.

\subsubsection{Connectivity Features (CF) } Beyond analyzing individual brain regions, CF offers a powerful tool for understanding the interplay between different brain areas in EEG signal processing. These features aim to quantify the functional relationships and synchronization patterns between various regions, providing valuable insights into the coordinated activity underlying cognitive processes and brain function.  Coherence \cite{coh_CF} is a widely utilized CF technique that measures the degree of correlation between the activities of different brain regions in the frequency domain. Phase synchronization \cite{PS_CF} goes a step further, evaluating the temporal alignment of neural oscillations across different regions, which can reveal intricate patterns of neural communication pivotal for cognitive and motor functions. 

\subsection{Model Building}
Many researchers have dedicated their efforts to translating human brain signals into text, exploring various innovative techniques. This field of study encompasses different methodologies for data collection, including EEG, fMRI, and ECoG. Each method offers unique insights and approaches to understanding brain activity. Deep neural networks, particularly RNNs \cite{Dewave,EEG-To-Text} and Long Short-Term Memory (LSTM) \cite{J-CRNN-BCI} networks, have demonstrated significant potential in decoding EEG signals into text. The BART \cite{EEG-To-Text} model is also emerging as a noteworthy tool in this domain. Bidirectional Auto-Regressive Transformers (BART), known for its effectiveness in various natural language processing tasks, is being adapted to interpret and convert EEG signals into coherent text. This adaptation signifies a promising convergence of advanced neural network architectures and neuroscientific data, potentially leading to more accurate and efficient EEG-to-text conversion methodologies. In this article, the primary emphasis is on those papers that utilize EEG signals to generate text. This approach presents a fascinating challenge and holds significant potential for advancements in neuroscientific research and assistive technologies.

\subsubsection{\textbf{DeWave} \cite{Dewave} } 
\begin{figure*}[t]
    \centering
    \includegraphics[width=1\linewidth]{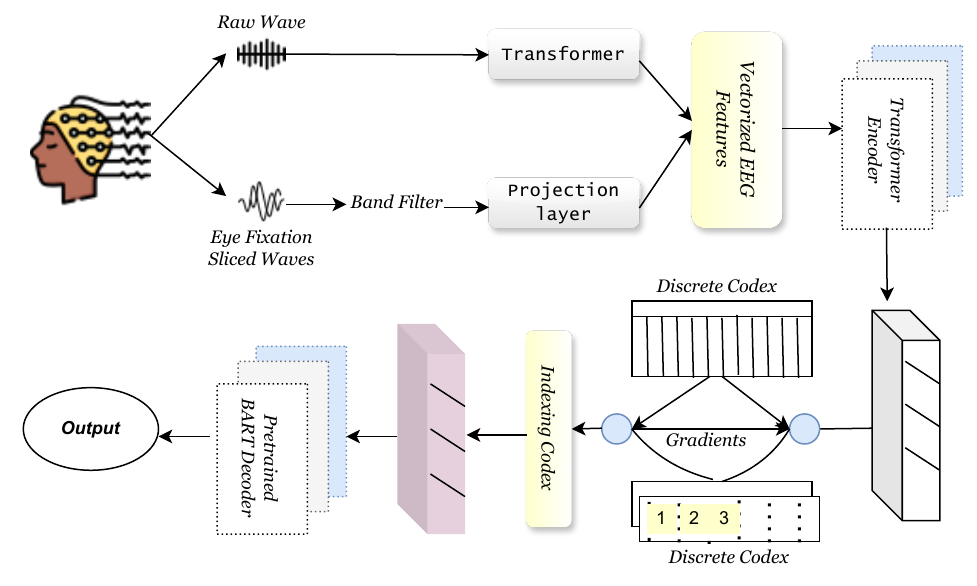}
    \caption{DeWave: A Transformer-Based EEG-to-Text Conversion Model.}
    \vspace{-10pt}
    \label{deWave}
\end{figure*}

DeWave is a novel end-to-end model for EEG-to-text translation using discrete codex encoding. It uses self-supervised learning and contrastive learning to establish the link between brain activity and language. DeWave eliminates the need for pre-processing steps like feature extraction, potentially simplifying the overall decoding process. Figure \ref{deWave} describes the architecture of the DeWave model.

Key steps in DeWave:

\textbf{Discrete Codex Encoding:} DeWave utilizes a technique called discrete codex encoding. This involves learning a codebook that maps continuous EEG signals into discrete code tokens. Imagine this codebook as a dictionary, where specific patterns in the EEG signal map to specific words or phrases.

\textbf{Self-supervised Wave Encoding:} The model learns this codebook through a self-supervised learning process. It essentially analyzes vast amounts of unlabeled EEG data and the corresponding text, allowing the model to discover the inherent relationships between brain activity and language.

\textbf{Contrastive Learning for Alignment:} DeWave employs contrastive learning to refine the alignment between the encoded EEG representations and the corresponding text. This process helps the model identify the most relevant EEG patterns that accurately reflect the intended text.

\textbf{Text Generation:} Finally, the model utilizes a decoder, often a Transformer-based architecture, which takes the encoded EEG representations (discrete code tokens) and translates them into the final text output, sentence by sentence.

\subsubsection{\textbf{MDADenseNet-AM} \cite{MDADenseNet-AM} } The MDADenseNet-AM model is a complex and sophisticated deep learning architecture designed for converting EEG signals into text. Here's a breakdown of its mathematical and operational structure:

\textbf{DenseNet Mechanism:} The model uses the DenseNet mechanism, where the output from the $m^{th}$ convolutional layer (denoted as $xd_m$) is determined by the application of 3×3 convolution filters $wd_dm$ on the outputs of all preceding layers. The mathematical representation of this output is given by:

\begin{equation}
xd_ m=\psi\left(\left[xd_0, xd_1, \ldots, xd_{m-1}\right]\right) \otimes wd_ m
\end{equation}

\textbf{Multiscale Dilated Convolution:} To address issues like aliasing, the model incorporates a multiscale dilated convolution operation. This operation allows the network to have a variable and adjustable receptive field, enabling it to capture information at different scales and contexts. The multiscale dilated convolution is characterized by different dilation factors, which determine the spacing of the convolution kernel elements and influence the channel obtained in the network.

\textbf{Integration with Attention Mechanism:} The attention mechanism is integrated into the MDADenseNet to focus on the most relevant information and suppress less important details. This is achieved by performing a weighted summation over the hidden layers of the network. Mathematically, for feature vectors $hh_v$, the environment vector $cc_v$ is calculated as follows:

\begin{equation}
cc_v=\sum_{v=1}^V aa_v hh_v
\end{equation}

Here, $aa_v$ represents the weights combined with the hidden state, indicating the importance of each feature in the context of the task.

\textbf{Parameter Optimization with EOWGMO:} The model is developed with an adaptive strategy for parameter optimization using the Eurasian Oystercatcher Wild Geese Migration Optimization (EOWGMO) algorithm. This approach optimizes parameters like optimizers and epochs in the DenseNet to enhance the accuracy and precision of the text conversion model. The optimization function can be represented as:

\begin{equation}
\begin{aligned}
& O N(2)=\arg \min \left\{O p D n t_{r r} E p D n t_{q r}\right\} \\
& \frac{1}{A c+P n}
\end{aligned}
\end{equation}

In this equation, $O p D n t_{r r}$ and $E p D n t_{q r}$ likely represent the ranges for the optimizer and epochs, respectively, while $A c$ and $P n$ are the accuracy and precision of the model.

\subsubsection{\textbf{EEG-to-Text} \cite{EEG-To-Text} } 
\begin{figure*}[t]
    \centering
    \includegraphics[width=1\linewidth]{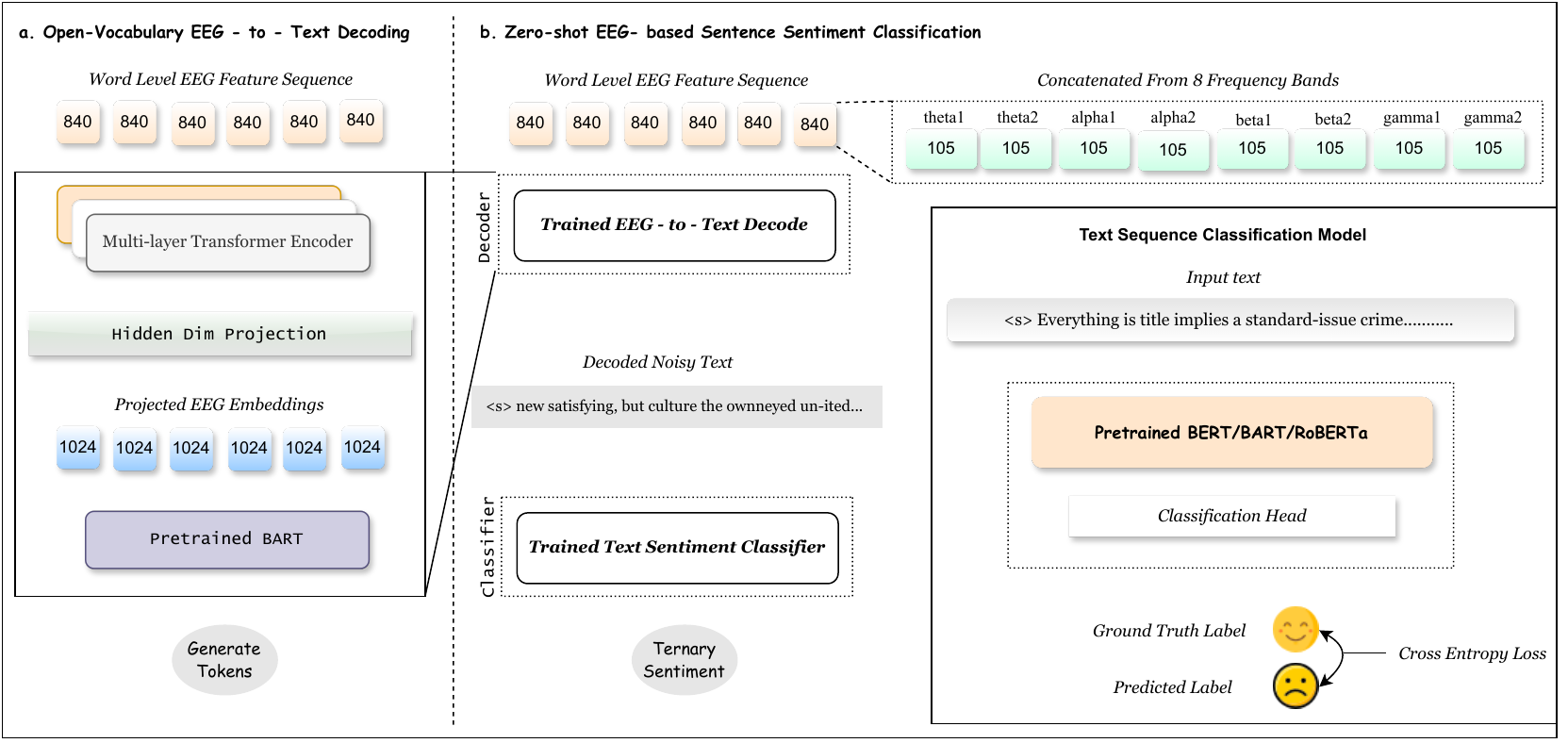}
    \caption{Integrative Framework for EEG Signal Decoding and Sentiment Analysis: Leveraging Pretrained Language Models for Text Generation and Emotion Classification.}
    \vspace{-10pt}
    \label{eeg-to-text}
\end{figure*}

This model, named EEG-To-Text, proposes a novel approach for open-vocabulary EEG-to-text decoding. It utilizes a combination of CNNs for feature extraction and LSTMs for sequence-to-sequence decoding. The model aims to overcome the limitations of pre-defined vocabularies and enable expressing a wider range of thoughts. The total working flow of the EEG-to-Text conversation model is shown in figure \ref{eeg-to-text}.

In the decoding stage of EEG-To-Text, LSTM networks play a crucial role in translating extracted EEG features into natural language sentences. Unlike traditional models, LSTMs excel at handling sequential data like EEG signals. Their internal memory mechanism allows them to retain information from previous time steps, capturing long-range dependencies between brain activity and intended meaning. Through a combination of "gates" that control information flow, the LSTM network analyzes current EEG features alongside previously predicted words, building the sentence word by word while considering the context of the entire sequence. This enables EEG-To-Text to decode diverse vocabulary and potentially capture the nuances of human thought.

\subsubsection{\textbf{J-CRNN-BCI} \cite{J-CRNN-BCI} } This paper proposed a deep learning framework for brain-computer interface (BCI) typing. This framework utilizes a joint convolutional recurrent neural network (J-CRNN) to decode motor imagery EEG (MI-EEG) signals, translating imagined typing movements into actual text output.

\textbf{Convolutional Neural Network (CNN):} The first stage of the J-CRNN acts as a feature extractor. It processes the raw EEG data, which captures electrical activity across the brain, and identifies spatiotemporal patterns related to specific imagined movements. By applying filters and performing convolutions, the CNN extracts relevant features that differentiate between different typing intentions.

\textbf{Recurrent Neural Network (RNN):} The second stage of the J-CRNN captures the sequential nature of the EEG signals. Unlike the CNN, which analyzes individual data points, the RNN considers the temporal dependencies between these points. This allows the model to understand the evolving patterns within the EEG signal over time, which is crucial for distinguishing between the unique sequences associated with each imagined letter.

By combining the strengths of CNNs and RNNs in the J-CRNN architecture, the model effectively extracts informative features from the EEG data while simultaneously capturing the temporal dynamics of brain activity. 

\section{Future Research Direction}
\subsection{Decoding complex thoughts and emotions} The recent advancements in EEG-to-text conversion are a significant achievement in neuroscientific research, offering new opportunities for communication, especially for those with speech and movement limitations. However, a notable gap in this domain is the current inability to accurately interpret and convey emotions through EEG signals. While researchers \cite{Dewave,EEG-To-Text} have developed models capable of translating brain activity into text, these models predominantly concentrate on the literal content and fail to consider the emotional context.  This limitation underscores the complexity of human emotions and their representation in brain activity, which existing technologies have not fully captured. It requires a deeper understanding of the neural correlates of emotions and the development of more sophisticated algorithms to address these challenges. Enhancing the emotional sensitivity of EEG-to-text systems will improve the fidelity of communication for users and has implications for fields like mental health. In this context, recognizing emotions can facilitate accurate diagnosis and effective therapy. Therefore, the next research phase in this domain focuses on integrating emotional intelligence into EEG-to-text systems, which will bridge the gap between technological capability and the nuanced spectrum of human expression.

\subsection{Data Source Diversification} The advancement of research in EEG-to-text generation is contingent upon the diversification of data sources, a crucial step towards enhancing the accuracy and applicability of these systems. Currently, EEG datasets used in research are often limited in size and diversity, leading to models that may exhibit poor generalizability across diverse populations or settings. To address this, it is important to integrate a broader array of data sources, encompassing datasets from diverse demographics, emotional states, and environmental situations.  Diversifying data sources improves the robustness of EEG-to-text models and ensures their inclusivity and adaptability in real-world scenarios. This approach aligns with the growing AI and machine learning trend towards creating more equitable and universally applicable technologies. Future research will likely focus on establishing large, varied datasets and developing models that can effectively learn from such complexity.  This will facilitate the creation of EEG-to-text systems that are precise, dependable, and reflective of the wide-ranging human encounter.

\subsection{Improving Accuracy and Reducing Errors} An important area of future research in EEG-to-text generation lies in enhancing accuracy and minimizing errors. This can be tackled through advancements in two key areas: deep learning architectures and signal processing techniques.  On the one hand, novel deep learning models specifically designed to handle the inherent noise present in EEG data can be explored. These models could use techniques like residual connections or attention mechanisms to extract more robust features and reduce the impact of noise on text generation.  On the other hand, developing advanced filtering and artifact removal methods can significantly improve the quality of EEG signals before they are fed into the text generation model. By integrating these approaches, researchers can significantly reduce errors and pave the way for more reliable and accurate communication through EEG-based text generation.

\subsection{Developed a Multi-model} Future EEG-to-text generation advancements can extend beyond textual outputs. Although present research mostly concentrates on utilizing EEG for precise text generation, there is potential for a broader exploration of multimodality. This might involve using a single EEG signal to generate text and create complementary representations of the user's intent. However, it's important to distinguish this from directly generating images or voices from the EEG data.  Instead, the focus would be on translating brain activity into additional modalities like simplified auditory representations of speech or basic visual icons that complement the textual output.  This would require significant breakthroughs in deciphering the complex neural correlates not just of language processing but also of sensory information processing within the brain. Overall, multimodal EEG-to-text generation presents a fascinating future direction, offering the potential for a richer and more nuanced communication experience.

\subsection{Cross-Domain Application} The expansion of EEG-to-text generation research into cross-domain applications represents a significant future direction, highlighting its potential beyond traditional boundaries. This approach involves applying EEG-to-text technologies across diverse domains such as healthcare, neuromarketing, and even the realm of creative arts, transcending its initial scope of aiding communication for those with disabilities. In healthcare, EEG-to-text systems could revolutionize patient care by providing non-verbal patients with a means to communicate their needs and symptoms. Moreover, in creative arts, this technology could offer a new medium for artists to translate their thoughts directly into textual form, pushing the boundaries of artistic expression. These cross-domain applications broaden the impact of EEG-to-text systems and encourage interdisciplinary collaboration, driving innovation and enhancing the depth of research in this field. To achieve successful cross-domain applications, it is important to modify the technology to align with the distinct demands and intricacies of each discipline. This task necessitates ongoing improvement and advancement of EEG-to-text systems.

\section{Conclusion}
The study of EEG-to-text translation represents a cutting-edge field in neuroscience and assistive technology that has demonstrated significant advancements in recent years. The progress in machine learning, namely in deep learning structures like RNNs, LSTMs, and transformer models, has established the foundation for advanced EEG decoding.  DeWave and similar innovations provide a glimpse of the possibility of creating more direct and efficient communication pathways for individuals with speech or physical difficulties. The non-invasive nature of EEG makes it a prospective option for wide-scale application, given its potential to provide a means of communication for individuals who are unable to speak or write traditionally.

Future research directions indicate the need to incorporate emotional intelligence into EEG-to-text systems, broaden the range of data sources, and improve the precision and fluency of these models. The potential to use this technology in multimodal communication and other fields indicates a significant impact that may exceed its existing limitations. This extension into cross-domain applications demonstrates the versatility of EEG-to-text systems and their capacity to enhance different facets of human existence, from healthcare to creative expression.

\bibliographystyle{IEEEtran}

\appendices

\end{document}